\documentclass[a4paper,fleqn,usenatbib]{mnras}

\usepackage{newtxtext,newtxmath}

\usepackage[T1]{fontenc}
\usepackage{ae,aecompl}

\usepackage[running]{lineno}

\usepackage{graphicx}	
\usepackage{amsmath}	
\usepackage{amssymb}	

\title[Abundance matching for low mass galaxies in the CDM and FDM models]{Abundance matching for low mass galaxies in the CDM and FDM models}

\author[P. Cristofari \& J.P. Ostriker]{
P. Cristofari$^{1}$ \& J.P.~Ostriker$^{1,2}$\thanks{E-mail: pc2781@columbia.edu}
\\
$^{1}$Department of Astronomy, Columbia University \\
$^{2}$Princeton University Observatory}

\date{Accepted XXX. Received YYY; in original form ZZZ}

\pubyear{2015}

\begin{document}
\label{firstpage}
\pagerange{\pageref{firstpage}--\pageref{lastpage}}
\maketitle

\begin{abstract}
Abundance matching between galaxies and halos has proven to be an informative technique, less dependent on detailed physical approximations than some other methods. We extend the discussion to the dwarf galaxies realm and to the study of the alternative hypothesis that the dark matter is composed of ultra--light particles: Fuzzy Dark Matter. We find that, given current observations, both CDM and FDM have difficulties with dwarf systems in the local group, but that, if mechanisms are proposed to alleviate these difficulties, they make radically different predictions for FDM and CDM at high redshift. These differences should allow clear observational tests with currently planned experiments, with the number of galaxies per unit volume with stellar mass over $10^6$ M$_{\odot}$ lower in FDM by factors of ($10^{-2.7},10^{-2.0},10^{-1.7},10^{-0.7},10^{-0.1}$) at $z=(10,8,6,4,2)$.

\end{abstract}



\section{Introduction}
The bulk of the matter of the Universe is believed to only interact through its gravitational effects. In  the current standard model, this dark matter (DM) is designated as Cold Dark Matter (CDM), a non--relativistic form of matter, and it can account for most of the cosmic observations on a wide range of scales and redshifts. 

Using CDM models, the community has significantly improved its understanding on the problem of galaxy formation and evolution~\citep[for reviews see][]{somerville2016,naab2017}. The history of galaxies is tightly connected to the history of DM halos hosting them, and therefore, the study of the DM halos -- and sub--halos -- structure has helped constrain the evolution of galaxies. 

On large scales, typically greater than several kpc, CDM models have been tested and can in general successfully match observations. However, on scales smaller than a few kpc, CDM models struggle to agree with observations. Several problems have been clearly identified. Amongst them, the famous problem of the over--prediction of the number of local group galaxies intermediate in mass between the two large systems (Milky Way, M31) and the numerous dwarf systems, often referred to as the \textit{too big to fail} problem  ~\citep[see e.g.][]{klypin1999,moore1999,boylan2011}. 
Very pragmatically, observations have clearly suggested that most of the Milky way satellite galaxies are hosted by DM halos of comparable dynamical masses within 300 pc~\citep{strigari2008}. Although efforts have been made to explain these observations~\citep[see e.g.][]{maccio2009}, this seems to be in tension with typical CDM models.

On the other hand, another alternative DM hypothesis has regained interest in the past decades, often referred to as~\textit{Fuzzy Dark Matter} or \textit{Ultra--light Dark Matter}~\citep{marsh2014,schive2014,schive2016}, considering the possibility of an axion--like, bosonic, DM particle, with a typical mass of the order of $\sim 10^{-22}$ eV. 
It is remarkable that this alternate DM candidate could dramatically affect the structures on Galactic scales, and therefore better account for small scale observations~\citep[see e.g.][and references therein]{hui2017}, without altering the successful predictions of CDM on large scales. 

 Other DM candidates have also been proposed such as \textit{warm dark matter} (WDM) and \textit{self--interacting dark matter} (SIDM). These candidates could also help alleviate the issues on the small scale, and though they will not be discussed in this paper, the reader can refer to~\citet[][and references therein]{anderhalden2013,lovell2016,schneider2017,murgia2017,read2017} for discussions on the WDM and SIDM hypotheses for local group galaxies.

The study of galaxy formation and their relationship with hosting dark matter halos can be done using very different approaches. Roughly, these approaches can be presented in three categories, even though many of them combine ingredients of the different categories~\citep[see e.g.][for a review]{baugh2006}.

First, gas--dynamical simulations, in which the gas dynamics is modeled with relatively few approximations regarding the physics of the baryons considered. These techniques are often computationally expensive, but can provide detailed descriptions of the internal structure of halos and galaxies~\citep{genel2014,hopkins2014,hopkins2017,hopkins2018,nori2018}. But these methods typically under resolve many important physical processes and so they must rely upon uncertain \textit{subgrid} recipes. 
Secondly, semi--analytical models, in which simplified descriptions are used to describe the different physical processes at stake. These methods usually rely on DM halo merger trees, obtained from N--body simulations of from Monte--Carlo methods~\citep{kauffmann1996,somerville1999,croton2006,somerville2015,henriques2015,lacey2016}. 

Thirdly, a category of methods often labelled \textit{semi--empirical} methods.
 These methods do not intend to describe the physical processes, but propose an effective model to link the DM halos and sub-(halos) to the hosted galaxies, with minimal physical assumptions.

A widely used approach of this category is the \textit{sub-halo abundance matching} (SHAM), described in~\citet{vale2004,vale2006} in detail and refined in several recent models~\citep{behroozi2010,moster2010,behroozi2013,moster2013,rodriguez2017,somerville2018}. In this paper, we will focus on this type of approach.
 
 These different methods all present advantages and disadvantages, intrinsically connected to the nature of the approach, and they often can complement each others. The SHAM technique is, by definition, the simplest type of approach. It relies on no detailed  physical modeling and only on the assumption that more massive halos tend to host more massive galaxies. 
 
  The effort required for the implementation of SHAM is minimal, but this technique however can be powerful for computing observable properties of DM halos and hosted galaxies.  Most of the successes of all these methods concern relatively massive systems, more massive than the Milky Way.
 
 Using SHAM techniques, we will investigate the question:~can CDM and FDM be in agreement with observations corresponding to low mass halos? We will especially focus on the case of low mass satellites of the Milky Way. In Sec.~\ref{sec:method} we quickly summarize the ideas supporting the SHAM techniques. In Sec.~\ref{sec:results} we present our results and the amendments needed for the SHAM approach to agree with observations of low mass galaxies, and we attempt to predict what forthcoming high redshift observations will tell us on the nature and abundance of low mass galaxies in the two dark matter models with extrapolations from observationally secure results at low redshift to falsifiable predictions at high redshift. Finally, in Sec.~\ref{sec:conclusions}, we summarize our results.

 \section{Method}
 \label{sec:method}
 The goal of the SHAM technique is to connect DM halos to their hosted galaxies. The idea at the core of this technique is that it is possible to statistically perform this connection, focussing on a given property of DM halos, such as for example their masses, and to then derive the relation between the stellar mass and the halo mass, the \textit{stellar--to--halo mass relation} (SHMR). The physical argument supporting this abundance matching technique is simply the realization that galaxies are formed in haloes through the accretion of gas, and that the expectation that the amount of gas is monotonically related to the mass of the halo. 
 The matching is performed as described in~\citet{vale2004,vale2006}, and one can then derive a relation between the stellar mass of a galaxy $M_{\star}$ and its hosting DM halo $M_{\rm h}$, following Eq.~\ref{eq:matching}.
 \begin{equation}
 \label{eq:matching}
 \int_{M_{\star}}^{+\infty}\phi(m) \;  \text{d}m =  \int_{M_{\rm h}}^{+\infty}n_{\rm h}(m_{\rm h}) \;  \text{d}m_{\rm h}, 
 \end{equation}
where $\phi$ is the galaxy stellar mass function and $n_{\rm h}$ the halo mass function. 
This assumes only that more massive halos host more massive galaxies.

It was originally proposed that the halo mass function should also take into account the number of sub--halos, especially for low masses haloes where the contribution of sub--haloes was shown to be potentially important. On the other hand, recent work~\citep{penarrubia2009,moster2010,reddick2013} has shown that the stellar mass of galaxies does not correlate easily with sub--haloes. This can be explained by the effects of tidal stripping, having a bigger effects on the sub--haloes than on the hosted star systems. 

Using this probabilistic approach, it is usual to define ${\cal P}(M_{\star}|M_{\rm h})$, the probability function that a halo of mass $M_{\rm h}$ hosts a galaxy of mass $M_{\star}$, as a log--normal distribution with a scatter $\sigma$ around the mean SHMR $\langle \text{log} M_{\star}(M_{\rm h}) \rangle$~\citep[see e.g.][for a clear introduction on the matter]{rodriguez2017}: 
 \begin{equation}
 {\cal P}(M_{\star}|M_{\rm h})= \frac{1}{\sqrt{2 \pi \sigma^2}} \times \text{exp} \left[- \frac{(\text{log} M_{\star} - \langle  \text{log} M_{\star} (M_{\rm h})\rangle )^2}{ 2 \sigma^2} \right]
 \label{eq:P}
\end{equation}

The value of $\sigma$ is not obvious. Firstly, because, although it can be constrained by observations, these observations are limited, and usually are not applied to DM halos below $\approx 10^{10} M_{\rm \odot}$. The range of DM halo mass below this value is therefore not well constrained by observations and potentially open to significant variations. Moreover, the meaning associated with $\sigma$ can vary. The scatter introduced by $\sigma$ in the matching accounts for a intrinsic possibility of some DM halos hosting galaxies of different masses, or in other words, for galaxies of a given mass to be embedded in DM halos of different masses. But in addition to this scatter, observables such as galactic stellar mass functions are subject to random errors, which can also be accounted for by modifying $\sigma$, i.e. including the dispersion from  observations in $\sigma$. Recent teams have been therefore able to account for these two effects writing $\sigma$ as~:
\begin{equation}
\sigma = \left( \sigma_{\rm h}^2 + \sigma_{\star}^2 \right)^{1/2} 
 \end{equation}
 where $\sigma_{\rm h}$ accounts for the intrinsic scatter in the halo--star matching, and $\sigma_{\rm \star}$ for the scatter due to random errors in the stellar mass observations. From observations, \citet{rodriguez2015} estimated~$\sigma_{\rm h}= 0.15\;  \text{dex}$ and~\cite{behroozi2010} estimated $\sigma_{\star} = 0.1 + 0.05 z \; \text{dex}$ where $z$ is the redshift. This typical values seem to be in agreement with recent observations.~\citep{tinker2017,somerville2018}
 
In order to derive the SMHR, using Eq.~\ref{eq:matching}, we first need to know the galactic stellar mass function $\phi$. Observations can usually be satisfyingly fitted using a double Schechter function:
\begin{equation}   
 \phi (m) \text{d}m = \left[ \phi_{1}^{\star} \left( \frac{m}{m_{\star}}\right)^{\alpha_1}+\phi_{2}^{\star} \left( \frac{m}{m_{\star}}\right)^{\alpha_2} \right] \text{exp}\left( -\frac{m}{m_{\star}}\right) \frac{\text{d}m}{m_{\star}}
 \end{equation}
where $\phi_{1}^{\star}, \phi_{2}^{\star}, \alpha_1, \alpha_2$ and $m_{\star}$ are parameters adjusted to observations. In this paper we take as reference the value proposed in~\citet{behroozi2013} for $z=0$, in Table~1 of~\citet{davidzon2017} for $2 \lesssim z \lesssim 6$ and in Table~C.1 of~\citet{grazian2015} for $6\lesssim z \lesssim 8$. 

\begin{figure}
\includegraphics[width=.5\textwidth]{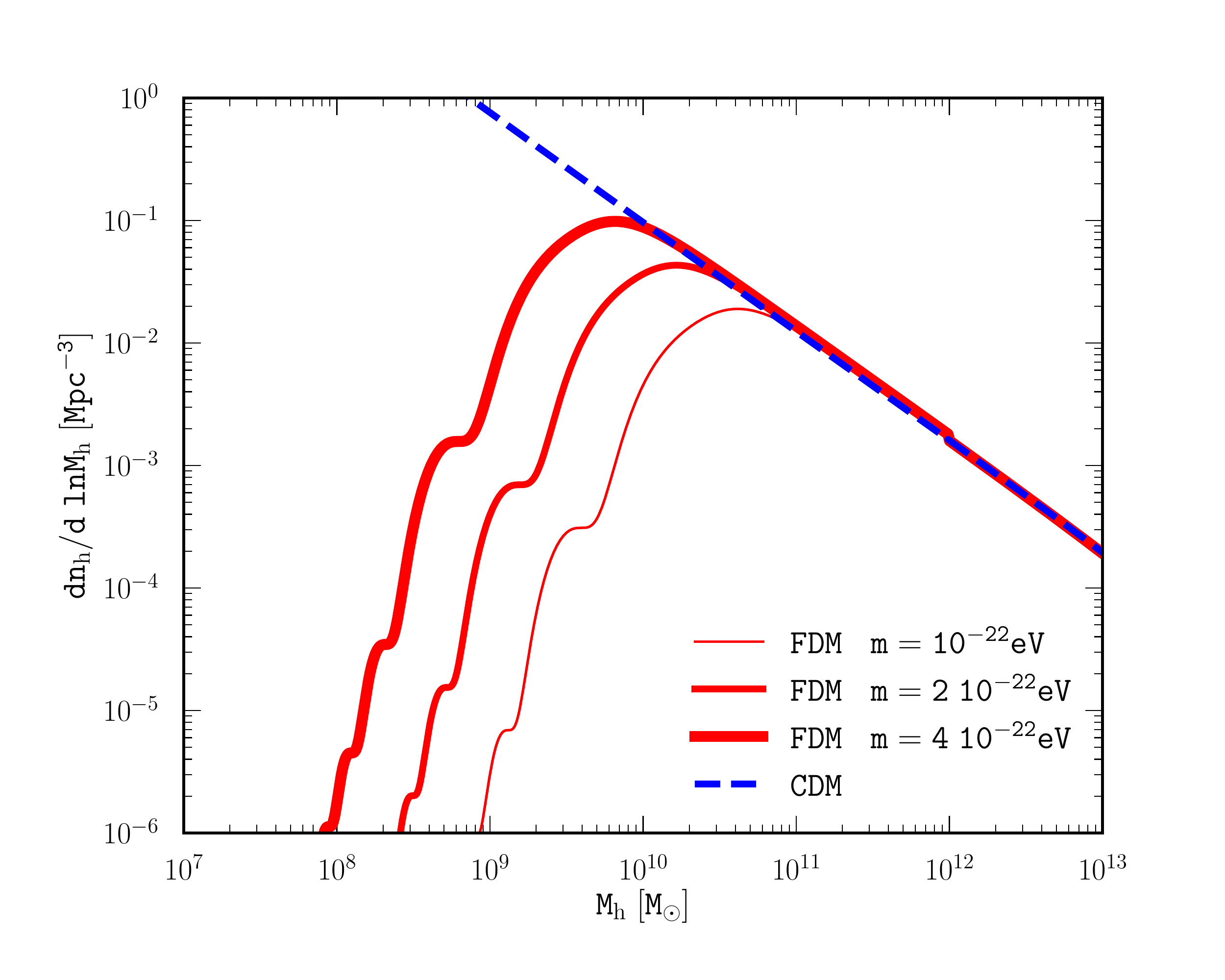}
\caption{Local (z=0) halo mass function in CDM model (blue dashed line) and in FDM model computed following the approach of~\citet{mihir} (red solid lines). From thin to thick line, the mass of the FDM candidate considered are $m$=1,2 and 4 $\times 10^{-22}$~eV.}
\label{fig:HMF}
\end{figure}

\begin{figure}
\includegraphics[width=.5\textwidth]{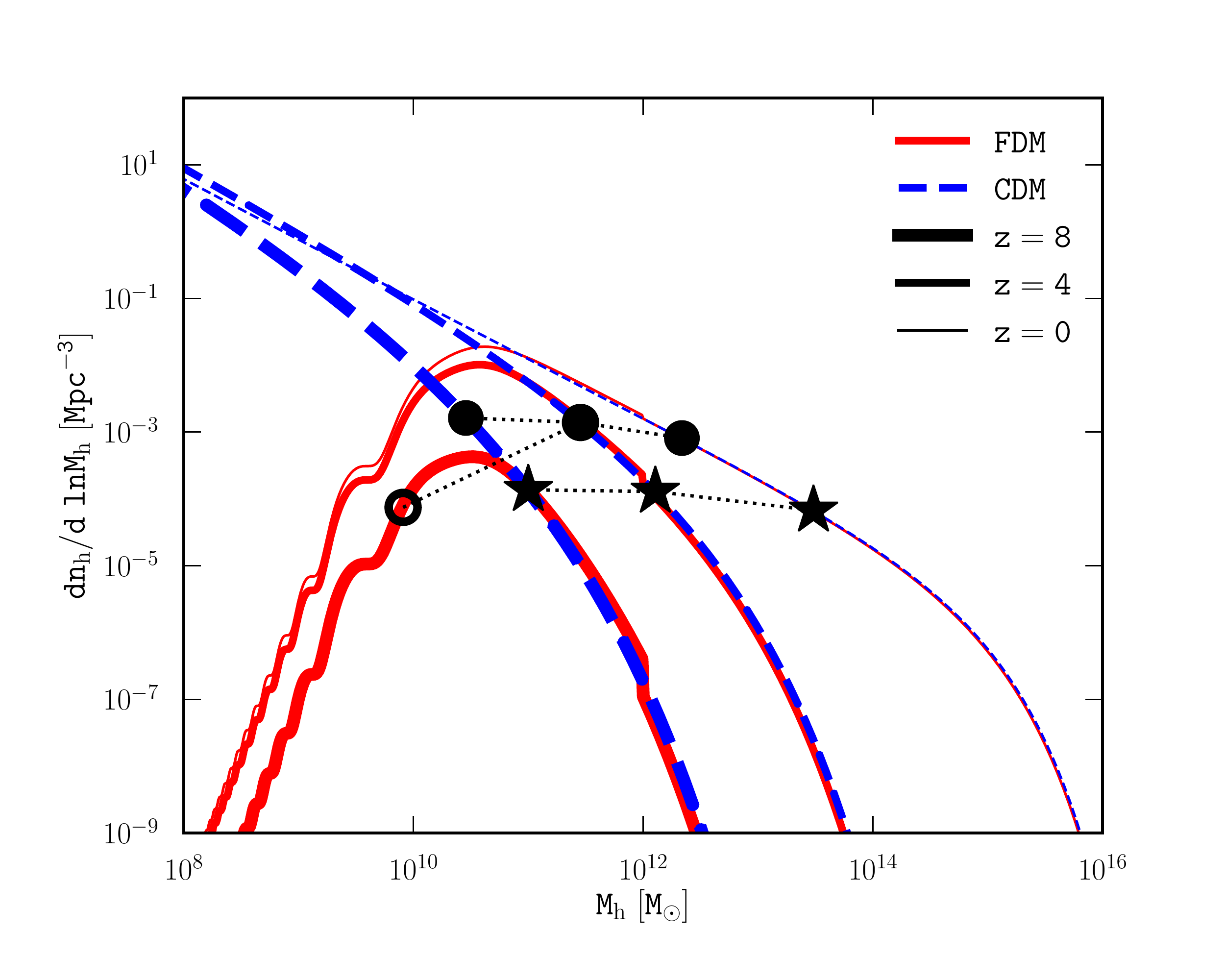}
\caption{Halo mass function in CDM (blue dashed line) and FDM  (red solid lines) models. The mass of the FDM candidate is $m=10^{-22}$ eV.  From thin to thick line, the redshifts considered are respectively $z=$0,4 and 8. The black stars represent the history of a halo of $10^{13.5}$ M$_{\odot}$ at z=0. The black circles represent the history of a halo of $10^{12.5}$ M$_{\odot}$ at z=0. The black unfilled circle represents the $10^{12.5}$ M$_{\odot}$ halo at $z=8$ in the FDM model.}
\label{fig:HMF1}
\end{figure}

The halo mass function $n_{\rm h}$ can be calculated using the Press--Schechter formalism and the linear power spectrum, which works well enough to fit numerical simulations for CDM~\citep{PS1974,bond1991,sheth2002}. Calculations in the case of FDM are a bit more subtle, because of the presence of the cut--off at low DM halo masses. Several calculations have been proposed~\citep{marsh2014,bozek2015,du2017} and a recent version using a sharp--k window function, that we use in this paper~\citep{mihir} is in good agreement with numerical simulations using initial FDM power spectrum~\citep{schive2016}.
As an illustration, we plot in Fig.~\ref{fig:HMF}  the halo mass function  in the case of FDM and CDM. We represent $\text{d}n_{\rm h}/\text{d}M_{\rm h}$ for  different masses of the FDM candidate m=1,2 and 4 $\times 10^{-22}$~eV. Note that in contrast to CDM, which has a steeply rising mass function for ever smaller subhalos, there is a peak in FDM and low mass haloes are rare. In Fig.~\ref{fig:HMF1}, halo mass functions in the FDM and CDM models are represented for redshift $z=0$, 4 and 8. The history of two halos of mass $10^{12.5}$ and $10^{13.5}$  M$_{\odot}$ at $z=0$ is represented at $z=(0,4$ and $8)$. 

\begin{figure}
\includegraphics[width=.5\textwidth]{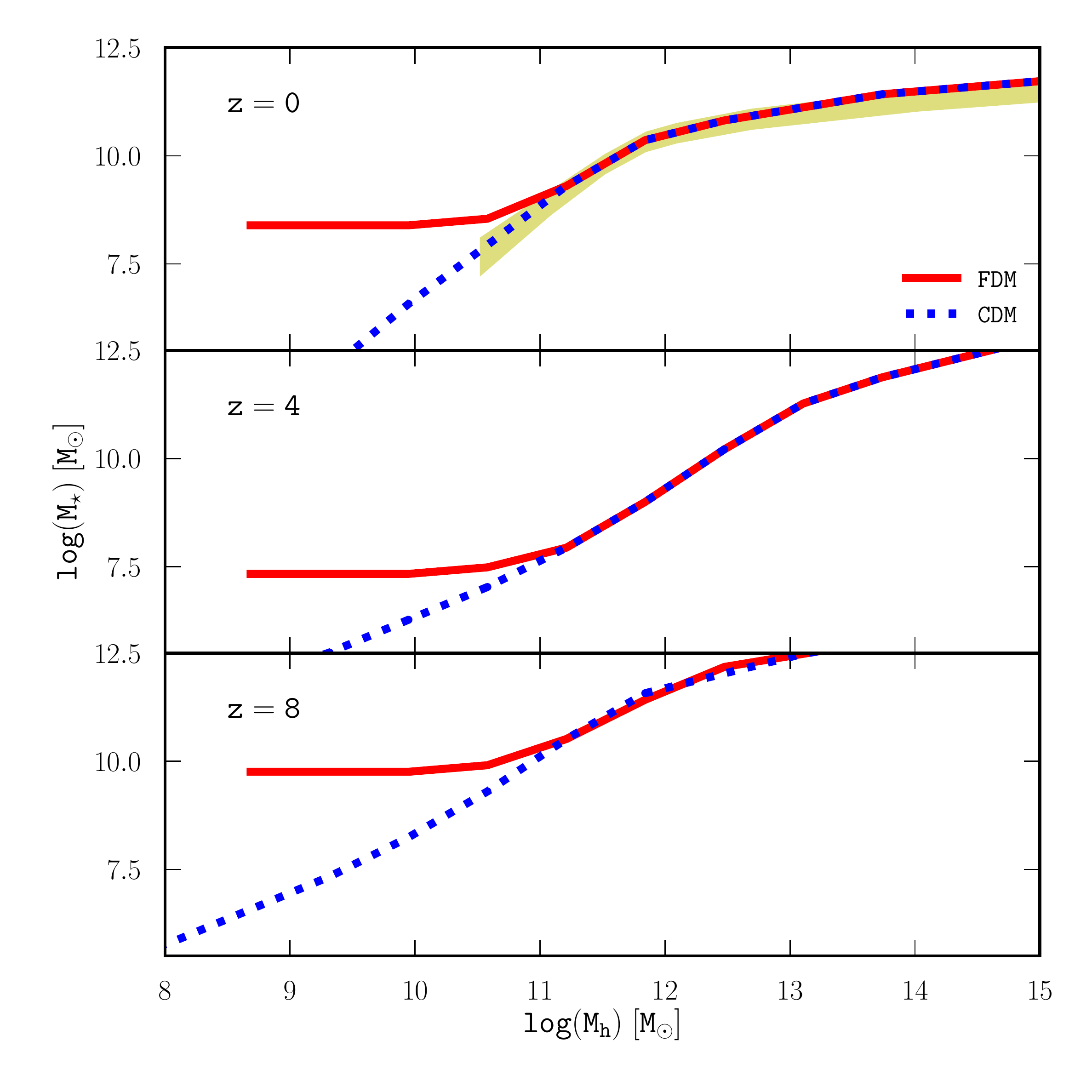}
\caption{Stellar mass--to--halo mass relation in the case of CDM (blue dashed lines) and FDM (red solid lines). The SHMR is shown for redshifts z=0,4 and 8 from top to bottom panel. The green shaded area in the top panel corresponds to the range of values (68\% confidence interval) derived by different groups from observations at redshift z=0, as summarized in Fig.~14 of~\citet{behroozi2013}. We terminate the FDM halo mass scale when the abundance of halo falls below $10^{-9}$ Mpc$^{-3}$. }
\label{fig:SHMR}
\end{figure}

Using Eq.~\ref{eq:matching}, the SHMR can be obtained. In Fig.~\ref{fig:SHMR}, as an example, we compute the mean SHMR in the case of CDM (blue dashed lines) and of FDM (red solid lines) for redshift 0, 3 and 6 (thick to thin lines), and a particle mass of $m=10^{-22}$eV. The green shaded area corresponds to typical SHMR derived from observations, representing the 68\% confidence interval of the values obtained by~\citet{moster2013,behroozi2013}, typically in agreement with other observations at redshift z=0~\citep{hansen2009,guo2010,yang2012,reddick2013} and with recent work~\citep{desmond2017,kulier2018,smercina2018}. SHMR obtained from observations usually concern halo masses $\approx 10^{10}-10^{12} \text{M}_{\odot}$. In the zeroth order, we see an overall agreement for masses greater than $\approx 10^{10.5} \text{M}_{\odot}$. Below this value, observations do not  constrain the SHMR efficiently,  although recent work discuss this matter~\citep[see e.g.][]{vanuitert2016,katz2017,read2017}. In our simple approach, the matching leads to remarkably different situations in the case of FDM and CDM. In the case of CDM, the slope of the SHMR obtained in the range $\sim 10^{11}-10^{12} \text{M}_{\odot}$ is conserved for smaller halo masses. In the case of FDM, the SHMR becomes brutally constant for halo masses below $\approx 10^{10} \text{M}_{\odot}$. 

This is directly due to the shape of the halo mass function  $n_{\rm h}$ in the case of FDM, as shown in Fig.~\ref{fig:HMF}: as noted for low masses, the number of halos decreases with the DM halo mass $M_{\rm h}$. Therefore, when performing the matching, below a certain mass, halos from all masses are associated to an essentially fixed galaxy mass, simply because very few low--mass halos are predicted under the FDM hypothesis.

  \section{Results}
 \label{sec:results}
 The SHAM technique presented briefly in Sec.~\ref{sec:method} can be used to investigate the number of galaxies corresponding to a given galaxy mass and halo mass. Indeed the SHMR illustrated in Fig.~\ref{fig:SHMR} does not account for the number density of galaxies, but only the relation between the two parameters ($M_{\star}$,$M_{\rm h}$). In the following, we investigate the number of objects expected in the  $M_{\star}$--$M_{\rm h}$ plan, and represent the relative numbers of expected objects on contour plots, as in Fig.~\ref{fig:contour}. We especially focus on the low range of $M_{\rm h}$ and $M_{\star}$, as that is where the discrepancies between models and observations are greatest. We take as a guide the data presented in~\citet{strigari2008}, where the authors studied the Milky Way satellite galaxies and found that several orders of magnitude in the inferred stellar mass were associated with a common mass within  300~parsec of $\approx 10^{7} \text{M}_{\odot}$, thus corresponding to a narrow range in DM halo masses ($\approx 10^{9.3}\text{M}_{\odot}$). 
 
 In the case of CDM, the halo mass can be estimated from the mass within 300~pc $M_{300}$~\citep{bullock2001,strigari2008}, using:
 \begin{equation}
 M_{\rm h} \approx 10^{9} \text{M}_{\odot} \left(\frac{M_{300}}{10^7 \text{M}_{\odot}}\right)^{2.8}
 \label{eq:Mh1}
 \end{equation}
 
Let us mention that Eq.~\ref{eq:Mh1} is used for simplicity to estimate the DM halo masses, and is subject to limitations. Indeed, deriving the mass density profile from the estimated mass within 300 pc naturally introduces uncertainties~\citep{maccio2009,errani2018}, and a definitive understanding of the different physical processes, such as tidal stripping, and their effects on the dynamical mass of satellites is still missing~\citep{read2018}.

\begin{figure*}
\begin{center}
\minipage{0.48\textwidth}
 \includegraphics[scale=0.35]{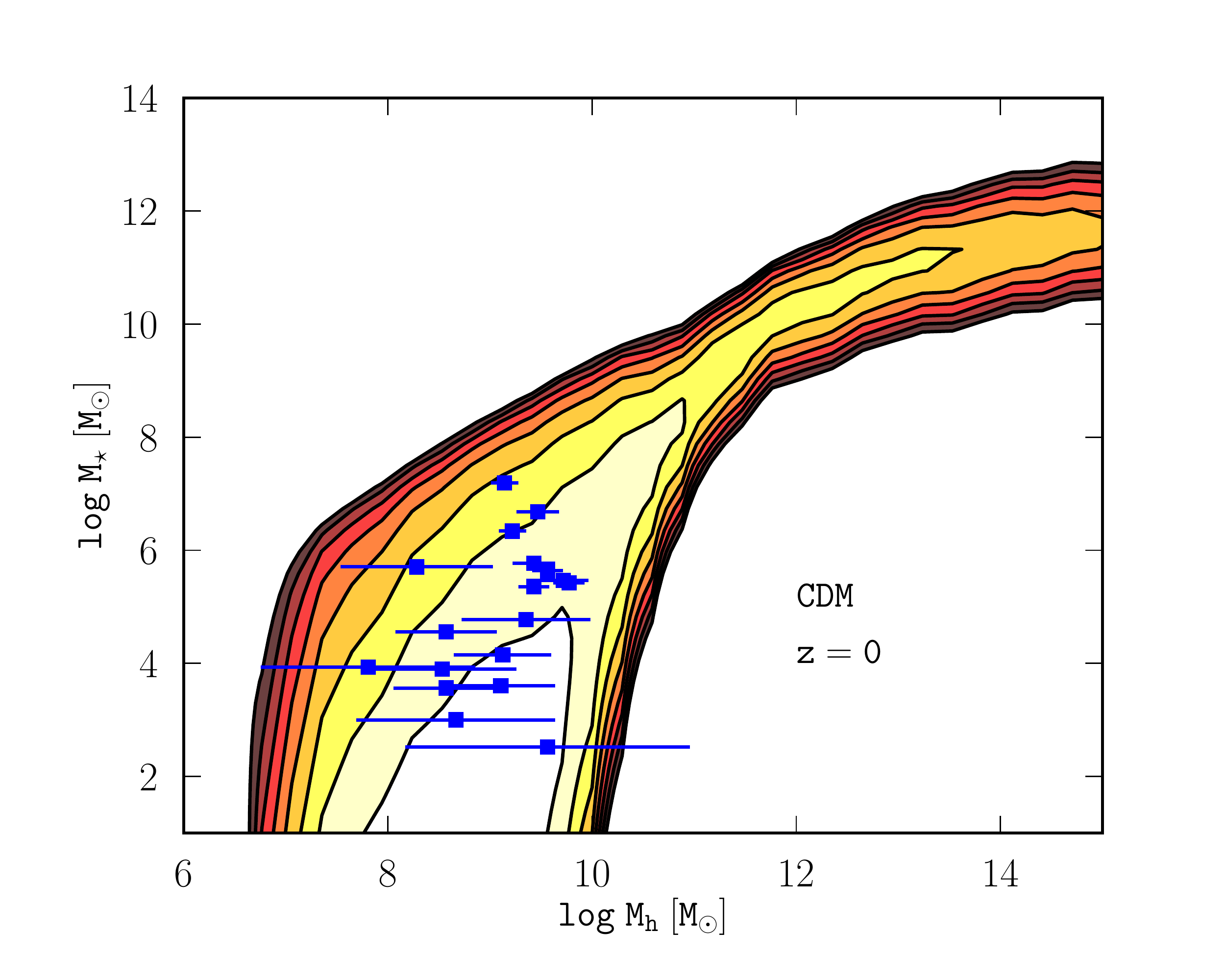}
\endminipage\hfill
\minipage{0.48\textwidth}
\includegraphics[scale=0.35]{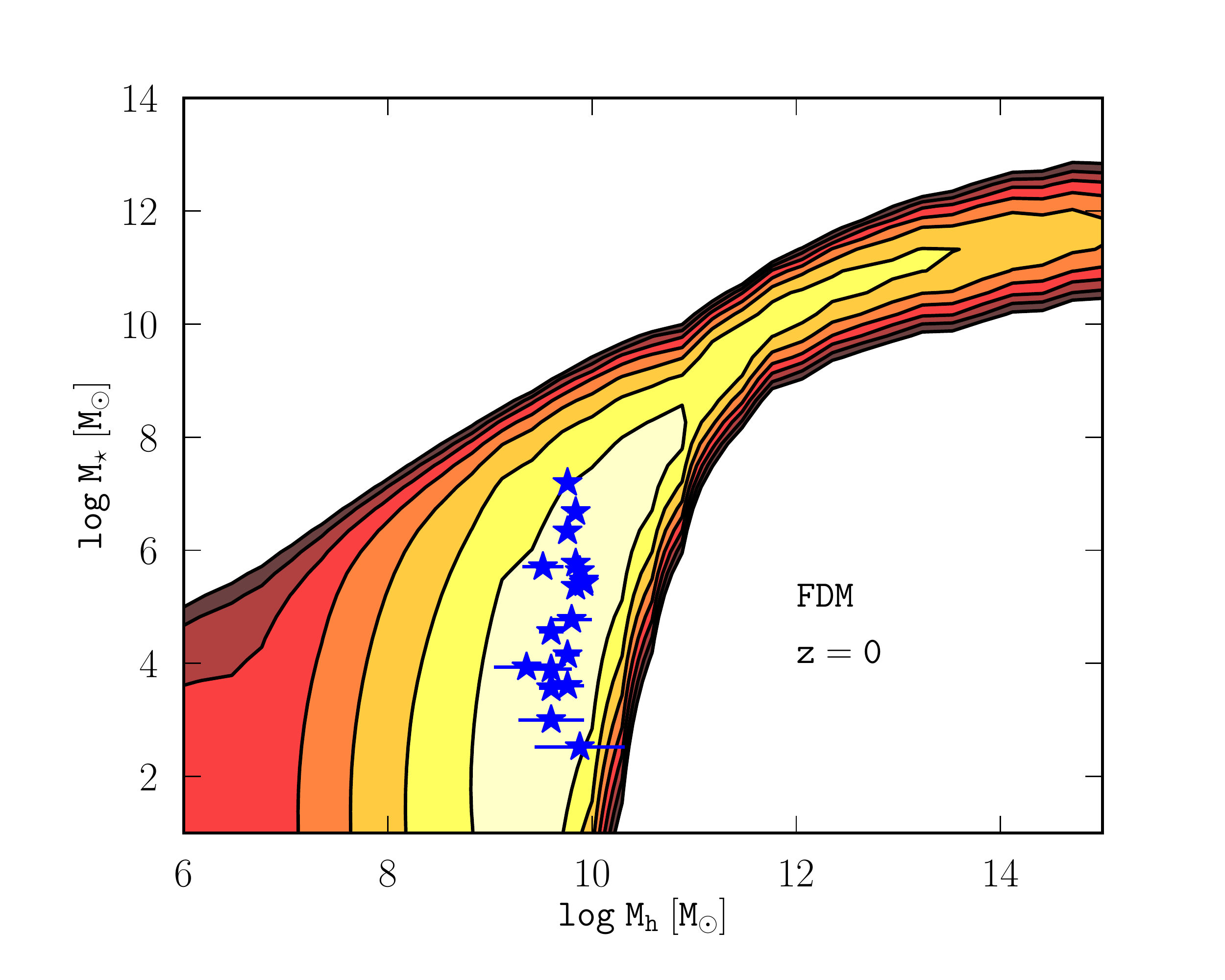}
\endminipage
\caption{Relative number of galaxies associated to the SHMR in the case of CDM (left panel) and FDM (right panel), for a redshift $z=0$. The number of expected galaxies is shown as contours, each contour corresponding to 2 orders of magnitude, with the highest central contour corresponding to $\geq 1$ object per comoving Mpc$^{-3}$. The data are taken from~\citet{strigari2008} (blue squares). In the case of FDM (blue stars) the corresponding halo masses are estimated as described in the text. }
\label{fig:contour0}
\end{center}
\end{figure*}

 In the case of FDM, the $M_{\rm h}$ estimated from $M_{300}$ changes, because of the different density profile. Indeed, in the FDM hypothesis, it has been shown that a dense \textit{soliton} core is expected in high mass DM halos~\citep{schive2014,schive2014b,urena2017}. The central soliton gives the galaxy a dark matter core rather than the usual $(\propto r^{-1}$) NFW cusp, and the central density is a steep function of the total halo mass with $\rho_{\rm FDM} \propto M_{\rm h}^{4/3}$: low mass galaxies have therefore very little dark matter within them in this model. For the central part to form the DM halo, the density profile can be described as follows~\citep{calabrese2016}: 
 \begin{equation}
 \rho_{\rm FDM}(r)\approx \frac{1.9(10\times m_{22})^{-2} r_{\rm c}^{-4} }{\left(1 + 9.1\times 10^{-2}(r/r_{\rm c})^2\right)^8} 10^{9} \text{M}_{\odot} \text{kpc}^{-3}
 \end{equation}
 where $m_{22}$ is the mass of the FDM candidate in units of $10^{-22}$~eV and $r_{\rm c}$ is the characteristic core radius of the halo. Typically, the soliton extends to $\sim 3~r_{\rm c}$, before returning to a profile similar to the typical  CDM density profile.   
 In our case, we treat $r_{\rm c}$ as a parameter and adjust it so that the mass within 300~pc corresponds the the one measured by~\citep{strigari2008}. 
The typical halo mass $M_{\rm h}$ can then be estimated directly as a function of $r_{\rm c}$: 
\begin{equation}
M_{\rm h} \approx 10^{9} \left( \frac{r_{\rm c} m_{22}}{1.6} \right)^{-3} \text{M}_{\odot} 
\end{equation}
 with $r_{\rm c}$ in units of kpc~\citep[see e.g.][and references therein for a didactic introduction]{calabrese2016}.
 Fitting to the Strigari results, we obtain, for FDM, $r_{\rm c} \approx 0.5$ kpc and $M_{\rm h} \approx 10^{9.7}$~$\text{M}_{\odot}$ within the virial radii for the typical case of $10^{7}$ $\text{M}_{\odot}$  within 300 pc. 
 In the case of FDM, the DM halo masses are represented in the right panel of Fig.~\ref{fig:contour0}.
 
 In order to achieve a relatively good agreement with the low mass galaxy data, for \textit{both} cosmological models we must introduce a lognormal dispersion in the matching with increasing dispersion as a function of decreasing halo mass, following Eq.~\ref{eq:sigma}. The dispersion is taken to be asymmetrical, and the lognormal distribution is truncated for stellar masses greater than  $f_{\rm baryon} \times M_{\rm h}$,with $f_{\rm baryon}= \frac{\Omega_{\rm baryon}}{\Omega_{\rm DM}}$. To do this, an exponential cut--off is introduced in the density distribution of galaxies in the form  $\propto \text{exp} \left[- \left(\frac{f_{\rm baryon} M_{\rm h}}{M_{\star}}\right)^{\alpha}\right]$,  where we adopt $\alpha = 3$ and $f_{\rm baryon} \approx 0.16$.  The motivation for this is to match recent results of simulations~\citep[see e.g.][]{onorbe2015,robles2017}.

 The number of galaxies ${\cal N}$ can then be expressed as:
 \begin{equation}
{\cal N} (M_{\star}) \text{d}M_{\star} = {\cal P}(M_{\star}|M_{\rm h}) \times n_{\rm h} (M_{\rm h})\times \exp \left[- \left(\frac{f_{\rm baryon} M_{\rm h}}{M_{\star}}\right)^{\alpha}\right] \text{d} M_{\rm h}
 \end{equation} 
where the dispersion $\sigma$, found in ${\cal P}$ (Eq.~\ref{eq:P}), is taken: 
 \begin{equation}
\sigma = \left( \sigma_{\rm h}^2 + \sigma_{\star}^2 + \sigma_{\rm m}^2 \right)^{1/2}  
 \end{equation}
 with: 
 \begin{equation}
 \sigma_{\rm m}(m)= \frac{A}{1+m/M_{\rm lim}}
 \label{eq:sigma}
 \end{equation}
 where $A$ and $M_{\rm lim}$ are two parameters.
 In Fig.~\ref{fig:contour0}, the parameters adopted are $A=1.5$ and $M_{\rm lim}=10^{10} \text{M}_{\odot}$.
 On these plots, the different contours correspond to 2 orders of magnitude of difference in number of objects. Eq.~\ref{eq:sigma} is required by the apparently observed fact - for the local group dwarf systems - that a small range in total mass corresponds to a large range in light output~\citep[see e.g.][]{martin2014,oman2016,vandokkum2018}.
 
In addition, the CDM model requires of a cut--off for low halo masses, that we introduce as an exponential cut--off in galaxy production probability acting at a mass $M_{\rm low}$. In Fig.~\ref{fig:contour} left panel,  $M_{\rm low}= 10^{7.8} \text{M}_{\odot}$. The motivation for this is the belief supported by simulations~\citep{robles2017} that in low mass galaxies, with low escape velocities, supernova feedback effectively blows out baryonic gas and catastrophically reduces star formation.  At low masses, the effects of reionization can also be important~\citep{sawala2015}.

   \begin{figure*}
\begin{center}
\minipage{0.48\textwidth}
 \includegraphics[scale=0.32]{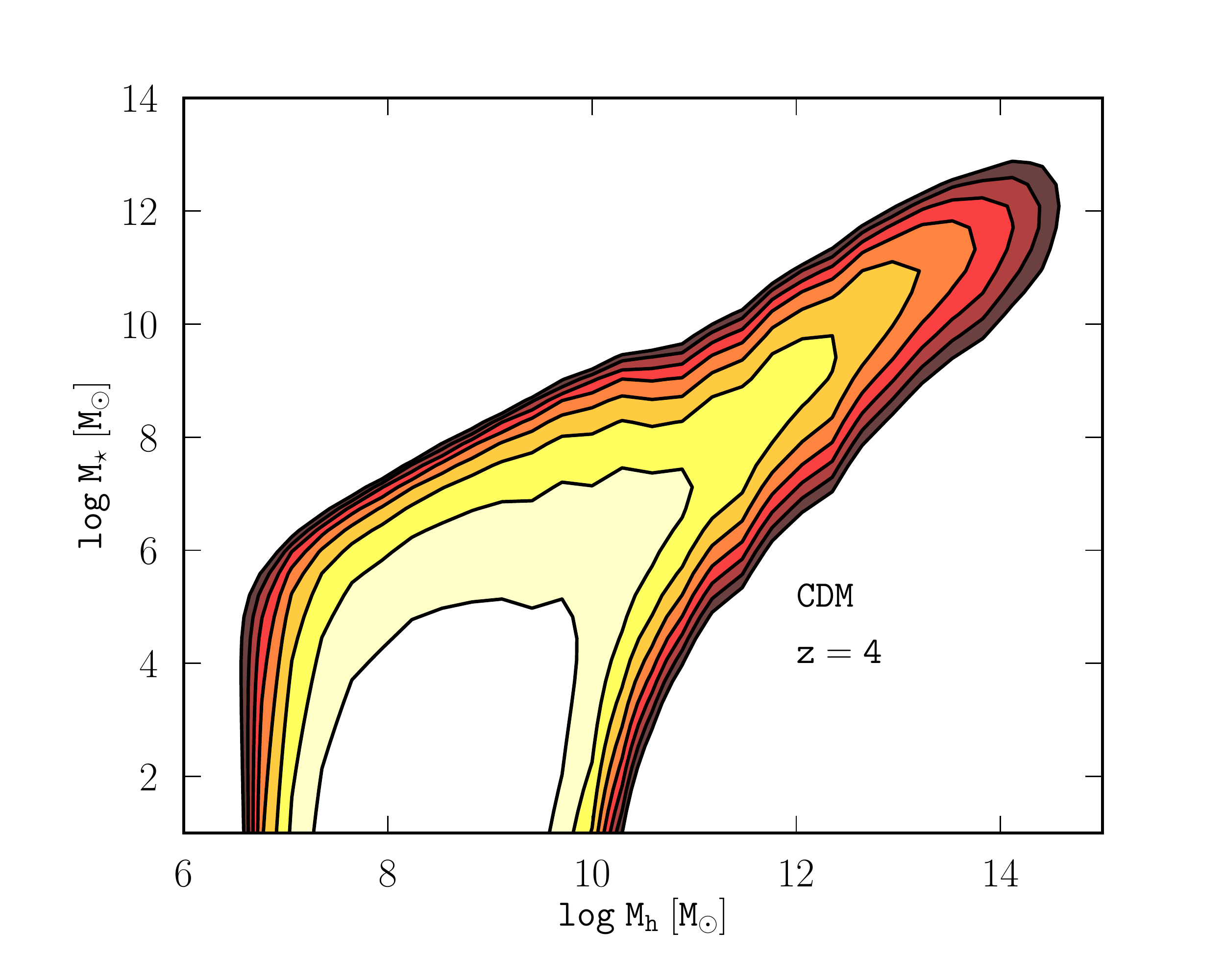}
\endminipage\hfill
\minipage{0.48\textwidth}
\includegraphics[scale=0.32]{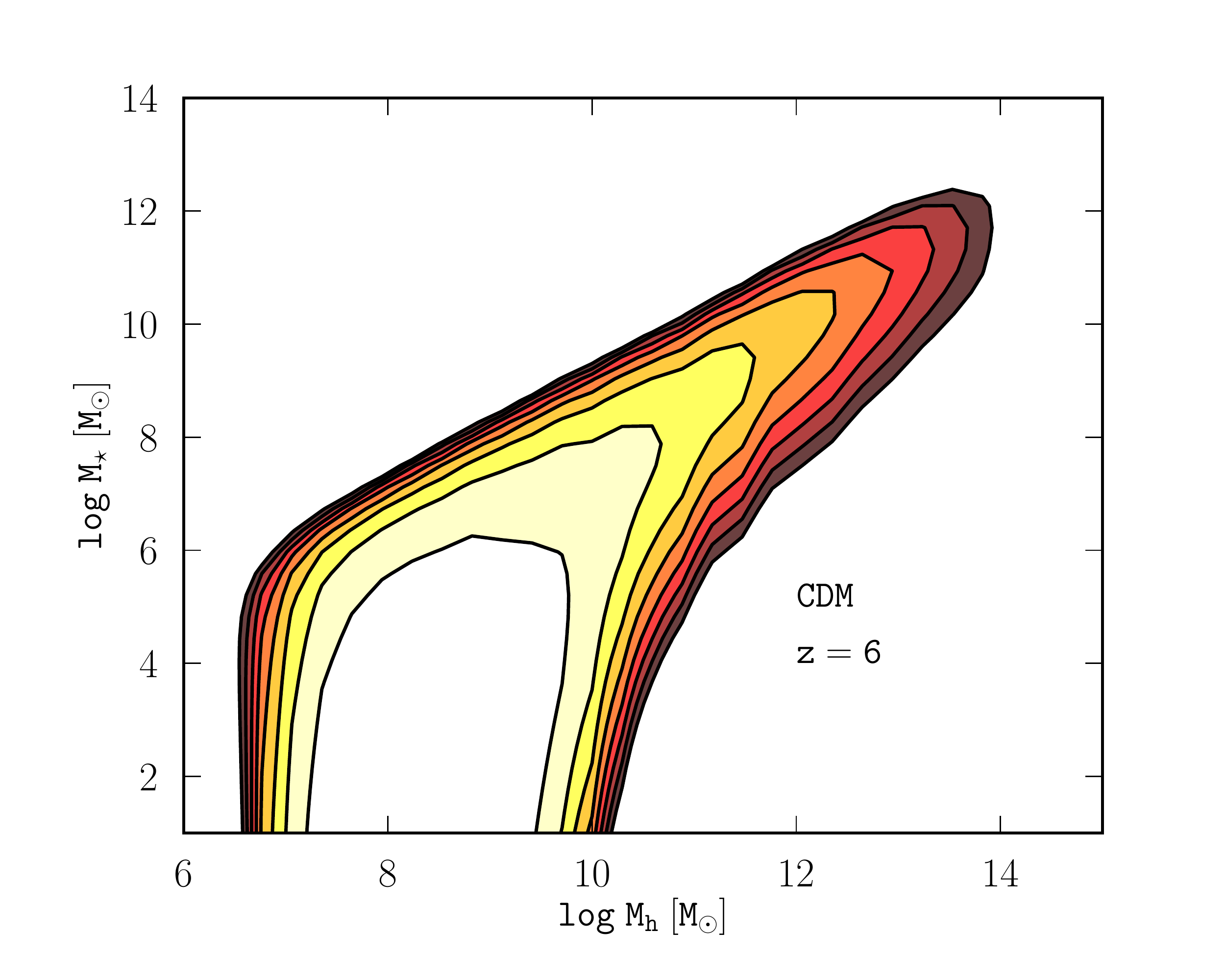}
\endminipage
\label{fig:contour}
\end{center}
\begin{center}
\minipage{0.48\textwidth}
 \includegraphics[scale=0.32]{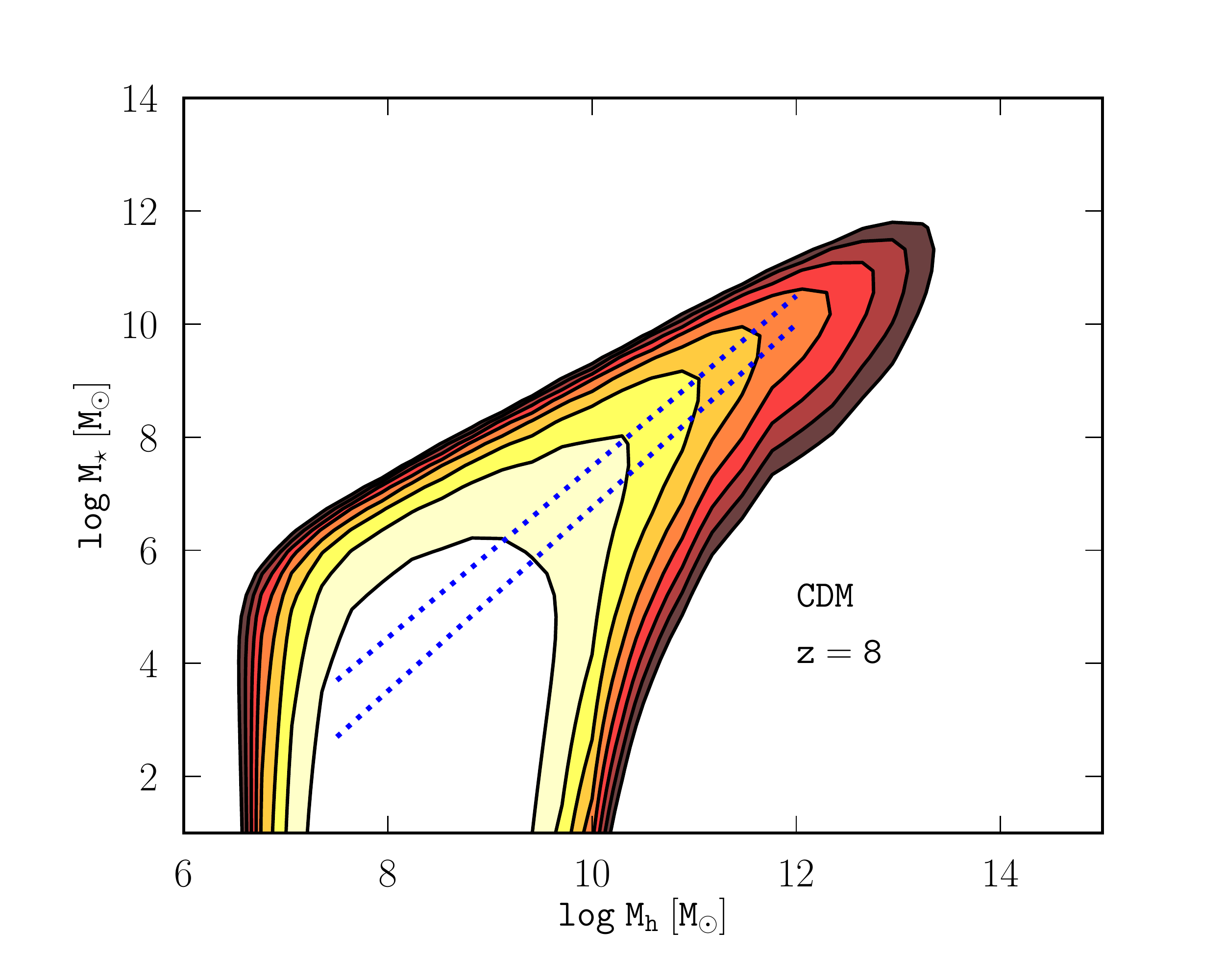}
\endminipage\hfill
\minipage{0.48\textwidth}
\includegraphics[scale=0.32]{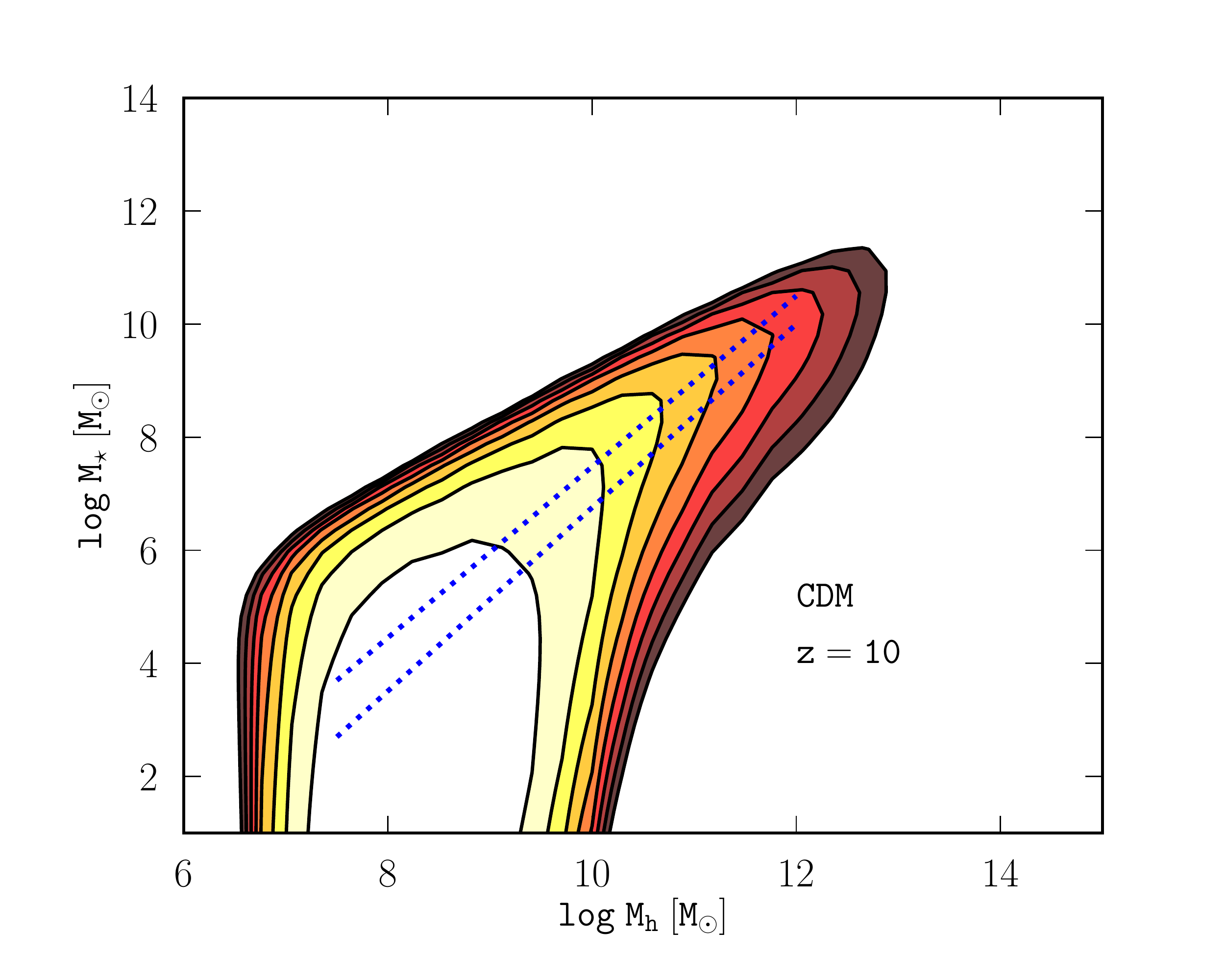}
\endminipage
\caption{Evolution of distribution of galaxies as a function of $z$ in the case of CDM, with same contours as in Fig.~\ref{fig:contour0}. At redshift $z=8,10$, the results of the FIRE simulations are shown (blue dotted lines, +/- 1 SD), taken from~\citet{ma2018}.}
\label{fig:evolutionCDM}
\end{center}
\end{figure*}
 In the case of FDM, no cut--off is required at low halo mass, because the halo mass function $n_{\rm h}$  declines for low $M_{\rm h}$. 
We only adjust the mass of the FDM candidate, taking $m = 1.9 \; 10^{-22}$~eV. 
Other options used to obtain agreements with the data are discussed in Sec.~\ref{sec:discussion}.
Finding a reasonable agreement with~\citet{strigari2008}, we then calculate the number of objects on the $M_{\star}-M_{\rm h}$ for different redshifts until $z=10$. In the case of CDM, the number of low mass galaxies in low masses halos remains somewhat unchanged as $z$ increases, and the evolution in $z$ mostly affects $M_{\star}\gtrsim 10^{10} \text{M}_{\odot}$. In the case of FDM, the evolution differs: as $z$ increases, at low and at high halo masses, the number of galaxy is suppressed. At $z=10$, only a small area of high density of galaxies exists near $M_{\rm h} \approx 10^{9.5} \text{M}_{\odot}$ and $M_{\star}= 10^{5}-10^{7} \text{M}_{\odot}$  remains~\citep{hirano2018}.

  \begin{figure*}
\begin{center}
\minipage{0.48\textwidth}
 \includegraphics[scale=0.32]{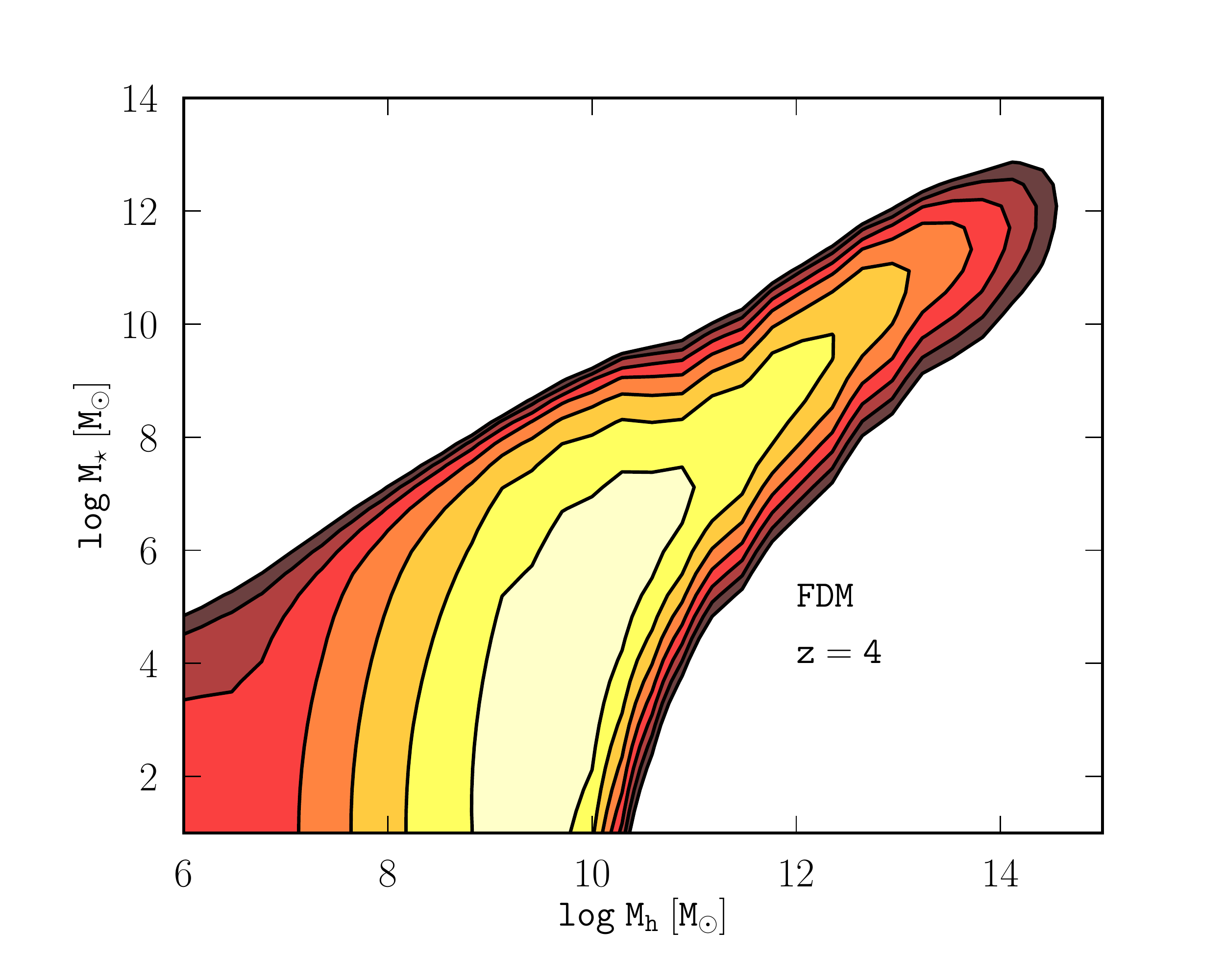}
\endminipage\hfill
\minipage{0.48\textwidth}
\includegraphics[scale=0.32]{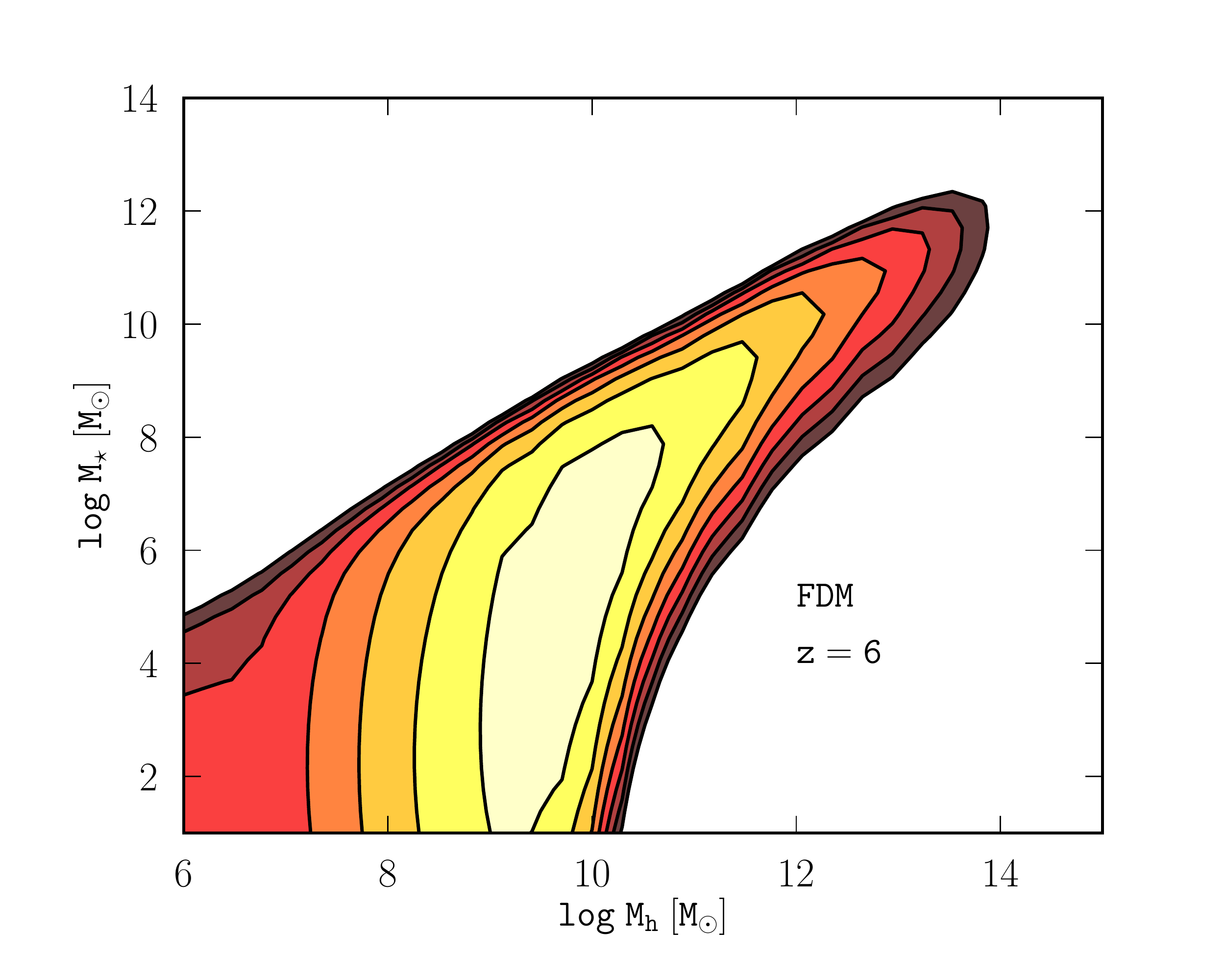}
\endminipage
\label{fig:contour}
\end{center}
\begin{center}
\minipage{0.48\textwidth}
 \includegraphics[scale=0.32]{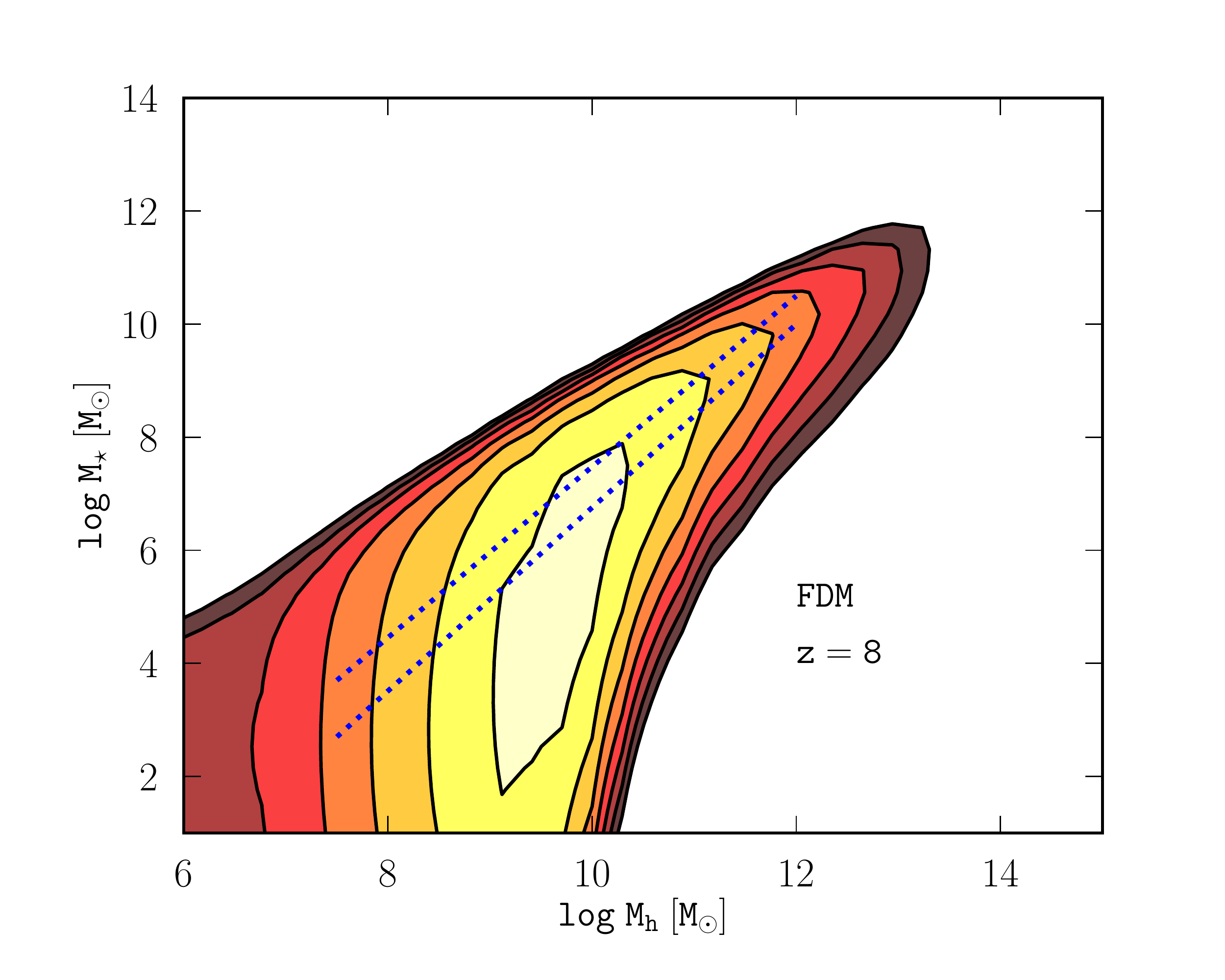}
\endminipage\hfill
\minipage{0.48\textwidth}
\includegraphics[scale=0.32]{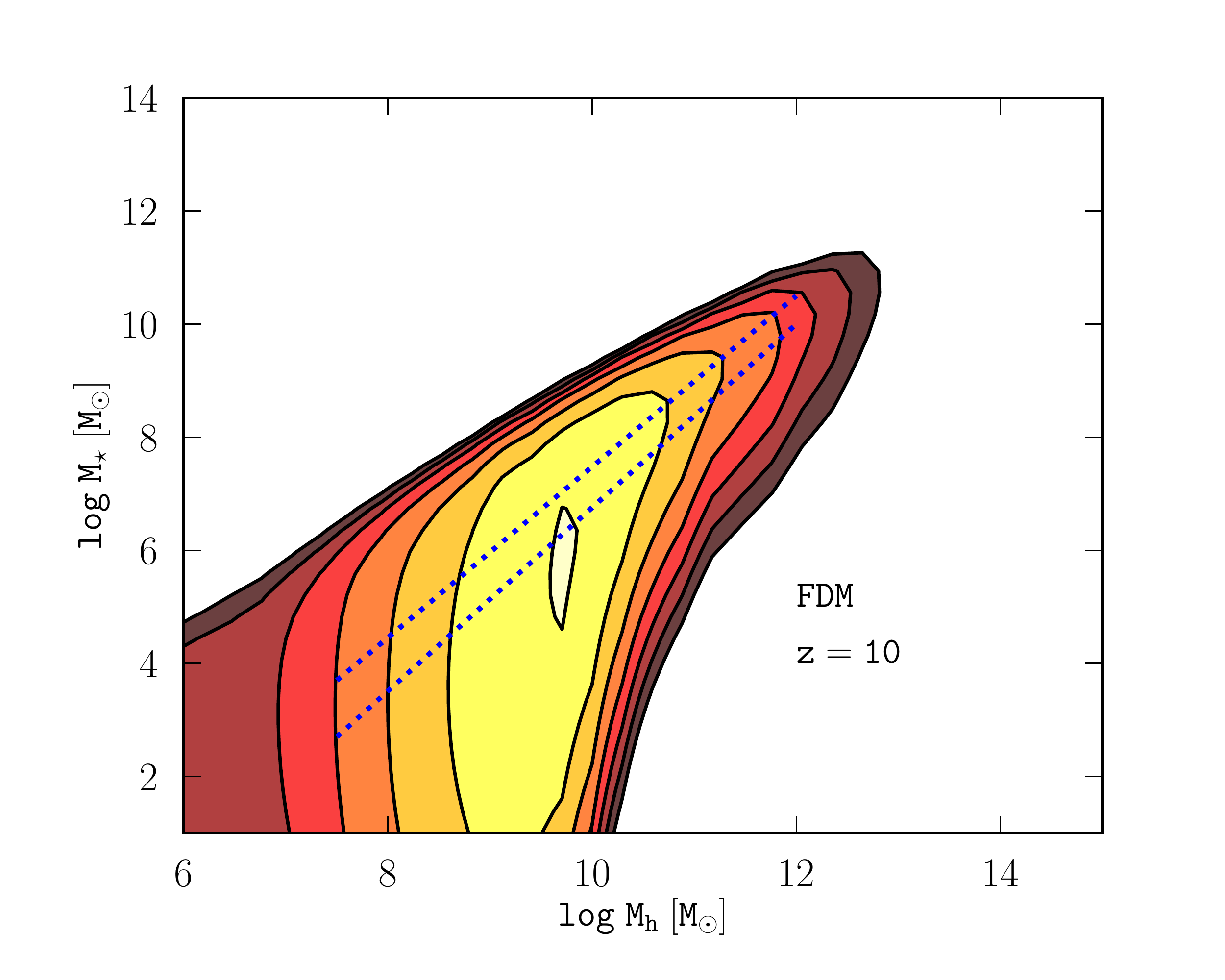}
\endminipage
\caption{Evolution of distribution of galaxies as a function of $z$ in the case of FDM, with same contours as in Fig.~\ref{fig:contour0}. At redshift $z=8,10$ the results of the FIRE simulations are shown (blue dotted line, +/- 1SD), taken from~\citet{ma2018}.}
\label{fig:evolutionFDM}
\end{center}
\end{figure*}

 \section{Discussion}
  \label{sec:discussion}
  The option chosen in Sec.~\ref{sec:results} in order to fit the low mass data consists in: 1) adopting a dispersion increasing with decreasing $M_{\rm h}$, 2) adding a low mass cut--off in the case of CDM, adjusting the mass of the DM particle in the case of FDM.

  Other physically motivated assumptions could reasonably be formulated in order to obtain a decent fit at $z=0$. For example, it is possible that a dispersion exists in the power spectrum used to calculate the halo mass function $n_{\rm h}$~\citep{mihir}. This would lead to a dispersion in $n_{\rm h}$. It is possible to model this dispersion in mass by simply convolving $n_{\rm h}$ with a log normal distribution. In the case of CDM, the effect is quite minimal because $n_{\rm h}$ ressembles a power--law. On the other hand, in the case of FDM, the dispersion flattens the low mass end of $n_{\rm h}$, an effect somewhat "analogous" to increasing the mass of the FDM candidate. To produce a plot similar to Fig.~\ref{fig:contour0} (right panel), one can for example assume $m=10^{-22}$~eV, and a dispersion of standard deviation $\mu \approx 0.3 \;\text{dex}$. The effect of this dispersion on $n_{\rm h}$ directly compensates the factor of $\approx 2$ in $m$.  
  
 It is also reasonable to assume that~\citet{strigari2008}'s velocity measurement need a correction in velocity due to contamination by unresolved binary stars~\citep{spencer2017}. To estimate that effect, let us consider the case in which we correct $V_{300}^2$ the squared velocity corresponding to the mass within 300~pc. As an illustration, we subtract $V_{\rm correct}^2= (4 \; \text{km/s})^2$ to $V_{300}^2$ to allow for binary star motions. In the case of CDM, this has the effect of displacing the data points to smaller values of $M_{\rm h}$, the greater displacement for the smaller values of $M_{\rm h}$: this corresponds, for the smallest $M_{\rm h}$ by $\sim$ half an order of magnitude. A decent fit is obtained for these adjusted values taking $M_{\rm LOW}=10^{7.5} \text{M}_{\odot}$ and $A=2$.  Overall, these new parameters do not affect substantially the evolution in $z$ presented in Fig.~\ref{fig:evolutionCDM}.
In the FDM hypothesis, the correction in velocity slightly decreases the "observed" values $M_{\rm h}$, the biggest effect remaining less than a factor of $\approx 2$. Keeping the parameters assumed in Sec.~\ref{sec:results} provides a decent fit to these corrected data. 

In this work, the changes of the baryonic ratio $f_{\rm baryon}$ with redshift $z$ are taken into account, parametrizing $f_{\rm baryon} (z) \propto (1+z)^{\beta}$ . We adopt $\beta=-0.2$, to match the results of recent work~\citep{moster2017}, which suggest that variations remain within a factor of $\approx 2$ for redshifts $z\leq10$.

Finally, we estimate the number of galaxies of stellar mass greater than $10^6$ M$_{\odot}$, $N_{\geq 10^6 \text{M}_{\odot}}$. At redshift $z=0$, the numbers for FDM and CDM models are  $ \approx 0.3$ Mpc$^{-3}$ and $\approx 0.5$ Mpc$^{-3}$ respectively. At redshift $z=10$, these numbers are $\approx 0.01$ Mpc$^{-3}$ and $\approx 4$ Mpc$^{-3}$ respectively for FDM and CDM, illustrating the divergence at high redshift.    
 At redshift $z= 8$ and $z=10$, results of the FIRE simulations are plotted~\citet{ma2018}. This illustrates that current simulations, performed in the context of CDM, do not reproduce the big dispersion in stellar mass at low halo masses that seem to be needed to match the observations. A study of such dispersion in the context of FDM simulations could provide valuable insight.

 Additional efforts have been made~\citep{rodriguez2012,rodriguez2013} to propose refined descriptions of the halo--mass--to--stellar--mass relation in the case of satellite galaxies, which could be added to our approach in a further study.

  \section{Conclusions}
 \label{sec:conclusions}
 We have investigated the capacity of SHAM techniques to provide a description of the SMHR compatible with observations. We especially study the low stellar mass range ($ M_{\star} \lesssim 10^7 \text{M}_{\odot}$) where tension with CDM models arises, and where it is not obvious how to match observations and models.  
 Using galaxy stellar mass functions fitted to observations and the calculated halo mass functions, we have in the CDM and FDM case, calculated the SHMR and derived parameters compatible with observations. 
 We found that SHAM techniques can reasonably satisfy for CDM models if allowance is made for a dispersion increasing with decreasing $M_{\rm h}$, and  cutting--off low galaxy masses below $M_{\rm LOW} \approx 10^{7.5} \text{M}_{\odot}$ for CDM. In the case of FDM, no cut--off is required, and considering a mass $m=1.9 \; 10^{-22}$~eV provides a reasonable concordance with observations. 
In agreement with data available at $z=0$, we have then qualitatively described the evolution with redshift of the population of galaxies in the $M_{\star}-M_{\rm h}$ plans. We illustrate that the differences in the CDM and FDM models are exacerbated as $z$ increases, leading to two remarkably different distributions of galaxies at $z \approx 10$, where the number of low mass galaxies is greatly suppressed in the case of FDM~\citep{schive2016}. In other words, galaxies will be much less abundant in FDM than in CDM at high redshift, but the ones that do exist will be more massive in FDM than they would be in CDM~\citep{leung2018}. For galaxies with stellar mass $\gtrsim 10^6$ M$_{\odot}$, the ratio of the number density of galaxies in FDM compared to CDM is $\approx (2\; 10^{-3}, 10^{-2}, 3\; 10^{-2}, 0.2,0.8)$ at $(z=10,8,6,4,2)$. 

In a future study, the case of isolated gas rich dwarfs~\citep{read2017} can be investigated using the method presented here. The study of these objects, for which the effects of tidal stripping and the quenching on the stellar mass are rather limited~\citep{jethwa2018,read2018}, could help refine the results presented here.

The detection, with current instruments, of globular cluster formation at redshifts $z\geq 3$~\citep{vanzella2017}  and of star--forming regions up to $z \approx 8$~\citep{bouwens2017}  illustrates the promising possibilities of next--generation instruments.

Future observations by next generation instruments, such as for example WFIRST~\citep{WFIRST}, LSST~\citep{LSST}, JWST~\citep{JWST,boylan2017,lovell2018} or Euclid~\citep{Euclid} will increase the statistics, especially at high redshift, and will help provide a critical test for the FDM hypothesis.

 \section*{Acknowlegment}
 The authors warmly thank P. Behroozi, G. Bryan, M. Kulkarni, S. Genel, K. Ichikawa, T. Naab, A. Rodriguez-Puebla, R. Somerville and the anonymous referee for helpful discussions and comments. 
 The authors acknowledge use of the Python package CosmoloPy~\citep{cosmolopy}. PC acknowledges support from the Frontiers of Science fellowship at Columbia University. 








\bsp	
\label{lastpage}
\end{document}